\documentclass[%
 reprint,
superscriptaddress,
 amsmath,amssymb,
 aps,
]{revtex4-2}

\usepackage{graphicx}
\usepackage{dcolumn}
\usepackage{bm}
\usepackage{xcolor}
\usepackage{hyperref}
\hypersetup{
    colorlinks=true,
    breaklinks=true,
    citecolor=black,
    linkcolor=black,
    menucolor=black,
    urlcolor=black,
}

\usepackage{siunitx}
\usepackage{amsmath}
\usepackage{orcidlink}
\usepackage{makecell}
\usepackage{tablefootnote}
\definecolor{orcidlogocol}{HTML}{A6CE39}   
\usepackage{lineno}
\usepackage{subfigure}
\begin{document}

\preprint{APS/123-QED}

\title{Mitigating effects of nonlinearities in homodyne quadrature interferometers}

\author{Johannes Lehmann\orcidlink{0009-0000-6998-4413}}
\affiliation{ Max Planck Institute for Gravitational Physics (Albert Einstein Institute), Callinstr. 38, Hannover, Germany}
\affiliation{ Institute for Gravitational Physics of the Leibniz Universit\"at Hannover, Callinstr. 38, Hannover, Germany}

\author{Artem Basalaev\orcidlink{0000-0001-5623-2853}}
\affiliation{ Max Planck Institute for Gravitational Physics (Albert Einstein Institute), Callinstr. 38, Hannover, Germany}
\affiliation{ Institute for Gravitational Physics of the Leibniz Universit\"at Hannover, Callinstr. 38, Hannover, Germany}

\author{Jonathan J. Carter\orcidlink{0000-0001-8845-0900}}
\affiliation{ Max Planck Institute for Gravitational Physics (Albert Einstein Institute), Callinstr. 38, Hannover, Germany}
\affiliation{ Institute for Gravitational Physics of the Leibniz Universit\"at Hannover, Callinstr. 38, Hannover, Germany}

\author{Sara Al-Kershi \orcidlink{ 0009-0004-1844-2381}}
\affiliation{ Max Planck Institute for Gravitational Physics (Albert Einstein Institute), Callinstr. 38, Hannover, Germany}
\affiliation{ Institute for Gravitational Physics of the Leibniz Universit\"at Hannover, Callinstr. 38, Hannover, Germany}

\author{Pascal Birckigt\orcidlink{0000-0001-8492-5964}}
\affiliation{ Fraunhofer Institute for Applied Optics and Precision Engineering, Albert-Einstein-Str.~7, Jena, Germany}

\author{Matteo Carlassara \orcidlink{0009-0007-2345-3706}}
\affiliation{ Max Planck Institute for Gravitational Physics (Albert Einstein Institute), Callinstr. 38, Hannover, Germany}
\affiliation{ Institute for Gravitational Physics of the Leibniz Universit\"at Hannover, Callinstr. 38, Hannover, Germany}

\author{Gabriella Chiarini\orcidlink{0009-0001-2977-5825}}
\affiliation{ Max Planck Institute for Gravitational Physics (Albert Einstein Institute), Callinstr. 38, Hannover, Germany}
\affiliation{ Institute for Gravitational Physics of the Leibniz Universit\"at Hannover, Callinstr. 38, Hannover, Germany}

\author{Firoz Khan \orcidlink{0000-0001-6176-853X}} 
\affiliation{ Max Planck Institute for Gravitational Physics (Albert Einstein Institute), Callinstr. 38, Hannover, Germany}
\affiliation{ Institute for Gravitational Physics of the Leibniz Universit\"at Hannover, Callinstr. 38, Hannover, Germany}

\author{Sina M. Koehlenbeck \orcidlink{0000-0002-3842-9051} }
\affiliation{E.\ L.\ Ginzton Laboratory, Stanford University, 348 Via Pueblo, Stanford, CA 94305, USA}

\author{Sarah L. Kranzhoff\orcidlink{0000-0003-3533-2059}}
\affiliation{ Universiteit Maastricht, Maastricht, 6200 MD, The Netherlands}
\affiliation{ Nikhef, Science Park 105, Amsterdam, 1098 XG, The Netherlands}
\author{Harald Lück \orcidlink{0000-0001-9350-4846}}
\affiliation{ Max Planck Institute for Gravitational Physics (Albert Einstein Institute), Callinstr. 38, Hannover, Germany}
\affiliation{ Institute for Gravitational Physics of the Leibniz Universit\"at Hannover, Callinstr. 38, Hannover, Germany}
\author{Pritam Sarkar\orcidlink{0009-0009-4054-6888}}
 \affiliation{ Max Planck Institute for Gravitational Physics (Albert Einstein Institute), Callinstr. 38, Hannover, Germany}
\affiliation{ Institute for Gravitational Physics of the Leibniz Universit\"at Hannover, Callinstr. 38, Hannover, Germany}
\author{Satoru Takano\orcidlink{0000-0002-1266-4555}}
 \affiliation{ Max Planck Institute for Gravitational Physics (Albert Einstein Institute), Callinstr. 38, Hannover, Germany}
\affiliation{ Institute for Gravitational Physics of the Leibniz Universit\"at Hannover, Callinstr. 38, Hannover, Germany}
 \author{Juliane von Wrangel\orcidlink{0009-0006-8431-0533}}
 \affiliation{ Max Planck Institute for Gravitational Physics (Albert Einstein Institute), Callinstr. 38, Hannover, Germany}
\affiliation{ Institute for Gravitational Physics of the Leibniz Universit\"at Hannover, Callinstr. 38, Hannover, Germany}
 \author{David S. Wu\orcidlink{0000-0003-2849-3751} }
\affiliation{ Max Planck Institute for Gravitational Physics (Albert Einstein Institute), Callinstr. 38, Hannover, Germany}
\affiliation{ Institute for Gravitational Physics of the Leibniz Universit\"at Hannover, Callinstr. 38, Hannover, Germany}

\date{\today}

\begin{abstract}
    Homodyne Quadrature interferometers (HoQI) are an interferometric displacement sensing scheme proven to have excellent noise performance, making them a strong candidate for sensing and control schemes in gravitational wave detector seismic isolation. 
    Like many interferometric schemes, HoQIs are prone to nonlinear effects when measuring displacements. 
    These nonlinearities, if left unsuppressed, would substantially limit the use cases of HoQIs.
    This paper first shows a means of measuring and quantifying nonlinearities using a working HoQI and a mechanical resonator. 
    We then demonstrate a method for real-time correction of these nonlinearities and several approaches for accurately calibrating the correction technique. 
    By correcting in real time, we remove one of the biggest obstacles to including HoQIs in upgrades to future gravitational wave detectors. 
    Finally, we discuss how to post correct data from HoQIs, suppressing even further the nonlinearity-induced errors, broadening the appeal of such sensors to other applications where measurement data can be reconstructed after the fact.  
    We demonstrate all of this on a working HoQI system and show the measured suppression of nonlinear effects from each of these methods. 
    Our work makes HoQIs a more broadly applicable tool for displacement sensing. 
\end{abstract}

\maketitle

\section{Introduction}
A wide range of fundamental physics experiments are sensitive to external vibrations which disturb our measurements~\cite{Richardson2020,Kang2006,Danzmann1996,Bronowicki2006,Matichard2015a,Matichard2015b}. 
Displacement sensors are used to monitor the relative motion between components of these experiments~\cite{Carbone2012,Cooper2018,Smetana2022} or motion of the experiment against an inertially suspended mass~\cite{Rodrigues2003,Christophe2015,MowLowry2019,Cooper2022,Hines2023,Heijningen2023,Carter2024b,Carter2025}. The latter provides a measure of absolute forces applied to the experiment.
Once the overall motion and forces are known, they can either be subtracted from measurements in post processing~\cite{Kang2006,Heijningen2023}, or used in active control loops. The control loops use mechanical actuators driven by sensor inputs to directly suppress motion-induced disturbances~\cite{Matichard2015a,Matichard2015b}. 
\par
The measurement of gravitational waves creates the most rigorous demands on vibrational isolation of any demonstrated experiment today. Gravitational wave observatories must isolate the motion of their test masses down to below the arm length changes induced by gravitational waves, currently with a target of the order of $10^{-20}\,\rm{m}/\sqrt{\rm{Hz}}$. For ground-based gravitational wave detectors such as LIGO~\cite{Aasi2015}, Virgo~\cite{Acernese2014} and KAGRA~\cite{Akutsu2021}, sophisticated passive suspension systems are used to isolate the test masses from ground motion in the frequency band they target for gravitational waves~\cite{Robertson2002,Accadia2011,Ushiba2021}. 
\par
The LIGO detectors supplement the passive suspensions with actively controlled isolation platforms~\cite{Matichard2015a,Matichard2015b}.
These platforms use a range of displacement and inertial sensors, combined in both feedback and feedforward control. Any error in the measurement of motion from the displacement and inertial sensors will be directly injected into the suspension system. Despite the sophistication of the platforms and their sensors, the residual noise they introduce is still a limiting factor in the detectors at low frequency~\cite{Capote2025}. The upcoming third-generation detectors, such as Einstein Telescope~\cite{Punturo2010} and Cosmic Explorer~\cite{Reitze2019} have even more stringent requirements on seismic isolation. In order to meet these requirements, a whole new suite of improved displacement and inertial sensors is needed~\cite{Saffarieh2025}.
\par
Mismeasurements from displacement sensors can originate from two sources: the sensor's stochastic noise and calibration errors. The largest source of calibration errors is the sensor's nonlinear behaviour. A perfectly linear displacement sensor will respond to a displacement with a proportional signal. A nonlinear sensor will respond to displacement with a signal defined by a nonlinear function. If one has sufficient knowledge of this nonlinear function, the signal can be inverted to again be proportional to displacement, making a linear sensor. The nonlinear function can be arbitrarily complex, so accurately modelling or measuring it for any system can be problematic.
Typically, the consequences of nonlinearities will get substantially worse as the magnitude of input displacement is increased. 
\par
Interferometric displacement sensors are the proposed displacement sensing method for the isolation platforms in the next-generation gravitational wave detectors. There are already several such schemes which show excellent noise floors \cite{Cooper2018,Yang2020,Smetana2022,Eckhardt2023,ChalathadkaSubrahmanya2025}, with sufficient performance for feedback control, where the measured signal is actively suppressed and therefore signal ranges are typically in the vicinity of the noise floor. Many of the interferometric schemes, however, are susceptible to nonlinearities which become significant if the motion is on the scale of interferometric fringes, which is of the order of one micrometre. Coincidentally, this is roughly the same magnitude of motion as typical seismically induced motion due to the so-called microseismic peak.
A means of quantifying and correcting nonlinear behaviour induced by motion on this scale is needed  for the feedforward control of the isolation platforms, readout of high mechanical quality (Q) factor accelerometers~\cite{Hines2023,Heijningen2023,Carter2024b,Carter2025}, or subtraction of disturbances in post processing~\cite{Kang2006}. For the first two applications, it is not sufficient to simply quantify the nonlinearities; they must be corrected in real-time measurements.
\par
Homodyne Quadrature Interferometers (HoQIs) are well studied for displacement sensing of moving test masses~\cite{Cooper2018}. The technique offers sensitivities in the sub \SI{100}{\femto\meter} regime, at frequencies from 1\,Hz and above.
They are being used in the sensing of inertial test masses in several applications such as Cartesian Inertial sensors~\cite{Cooper2022,Kranzhoff2022, Carter2025}, rotational sensors~\cite{Ross2023}, frequency references~\cite{DiFronzo2024}, and multiple degree of freedom suspended sensors~\cite{Nagano2025,Prokhorov2024}. They are also used as sensors for the active damping of multi-stage optics suspensions~\cite{Mitchell2025}.

HoQIs are modified Michelson interferometers which use polarising optics to create signals proportional to the sine and cosine of the optical phase. This allows to track the optical phase, or displacement of the test mass, over multiple free spectral ranges (FSRs). Due to the nonlinear nature of sine and cosine functions, any imperfection in their measurement on the different photodiodes (PDs) introduces nonlinear errors in the phase readout.
\par
Nonlinearities have been characterised and measured for similar displacement sensors. With sufficiently large detected motion, errors in the quadrature signals of the optical phase can be corrected through fitting techniques, leading to sub-nm nonlinearities~\cite{Wu1996}. Those techniques can also be applied to live data, for example by correcting the signals before digitisation~\cite{Eom2001}. More recent investigations were able to show how nonlinearities can arise due to unwanted reflections of optical components~\cite{Hu2017} or due to misalignment~\cite{Lin2024} and that segmented fitting of the quadrature signals with phase dependent parameters can further reduce nonlinearities.
Nonlinearities were observed in interferometric sensors with different readout techniques as well~\cite{Smetana2023}. 
\par
In this paper, we investigate the correction of nonlinearity in the HoQI measurement, both in post-processing and in real-time in our digital data acquisition system (Control and Data System, CDS~\cite{Bork2001}). 
In section~\ref{sec:exp_setup}, we show a means of quantifying nonlinearities in HoQIs using a mechanical resonator. 
In section \ref{sec:compensating_nonlinearities}, we demonstrate that the inherent nonlinearities of HoQIs can be corrected to such a point that, in real-time measurements, their sensitivities are improved by an order of magnitude. If the data is not needed in real time, it can be further corrected and we discuss methods to do this in section \ref{sec:postcorrect}. 
Our methods are all demonstrated on a working HoQI and we demonstrate the improved performance of the HoQI using these correction techniques.

\section{Testing Nonlinearities in Displacement Sensors}
\label{sec:exp_setup}
The investigations carried out here were done with an existing, noise characterised HoQI~\cite{Kranzhoff2022,Carter2025} at the Albert Einstein Institute (AEI) 10\,m Prototype facility~\cite{Gossler2010}. The functionality of HoQIs has been explained in previous literature \cite{Cooper2018}, but we highlight the key aspects relevant to this work.

\subsection{Reconstructing Phase signals in Quadrature Inteferometers}
\label{subsec:reconstructing phase signals}
\begin{figure}
    \centering
    \includegraphics[width=.48\textwidth]{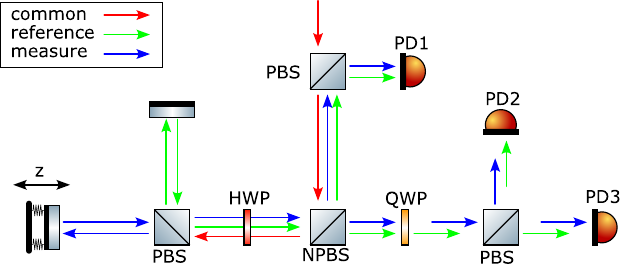}
    \caption{Beam paths in a HoQI. The beam is split by a polarising beam splitter (PBS) and the parts are sent to a reference mirror and measurement mirror. The beams containing phase information of the two different arms are shown with separate arrows despite being overlapped in reality. The returning light is split by a non-polarising beam splitter (NPBS) before being measured by 3 different PDs. The delay of one of the beams due to the quarter wave plate (QWP) is shown by a shift of the arrows.}
    \label{fig:hoqilayout}
\end{figure}

A HoQI features three PDs producing signals proportional to sine or cosine of the optical phase, which, in turn, is proportional to the length difference of two propagating lasers beams. Figure.~\ref{fig:hoqilayout} shows the beam paths through the different optics.
The beams typically have a common path up to an optic near the object whose motion we wish to measure, where they are split by a polarising beam splitter whose axis is aligned 45 degrees to the polarisation of the the incident light. The s polarised light is then sent to a reference mirror which is rigidly attached to the HoQI baseplate. The p polarised light is transmitted to a second mirror which is attached to the object whose motion we wish to measure, the test mass. The two beams therefore accumulate different optical phase, $\phi$, directly related to the motion of the moving object by
\begin{equation}
    \delta=\phi\frac{\lambda}{4\pi},
\end{equation}
where $\delta$ is the motion and $\lambda$ is the laser wavelength of the light.
The beams then return to the polarising beam splitter and are sent to the three PDs with different polarisation rotations applied to them. \par
In the ideal case, as detailed in Ref.~\cite{Cooper2018}, the different polarisation states create PD signals which can be combined to create two orthogonal quadratures:
\begin{equation}
P_1 - P_2 =  c \cdot \text{sin}(\phi) = Q_1, P_1 - P_3 =  c \cdot \text{cos}(\phi) = Q_2,
\end{equation}
where $P_n$ indicates the power on the respective PD, $c$ is the calibration constant, and $Q_m$ the respective quadrature.
As a result, if visualized on a $Q_1 Q_2$-plane, the data draws a circle, or a sector of a circle in case of sub-FSR motion. The phase can be recovered by taking a 2-argument arctangent of the quadratures:
\begin{equation}
\label{eq:arctan}
\phi = \text{arctan2}(Q_1,Q_2).
\end{equation}
\par
In case of motion over one FSR, the phase has to be unwrapped to correctly track the motion.

\subsection{Fused Silica Mechanical Resonators as a Means of Injecting Linear Motion}
Noise measurements of displacement sensors involve testing the sensor on a test mass rigidly fixed to the sensor, which would correspond to a constant $z$ in Fig.~\ref{fig:hoqilayout}. When nothing moves, the noise floor is simply the measured signal~\cite{Hines2020,Cooper2018,ChalathadkaSubrahmanya2025}. HoQI noise measurements have shown that noise floors below $10^{-13}$\,m/$\sqrt{\rm{Hz}}$ can be achieved for stationary test masses, but experiments have shown that these same noise floors are not always reached on dynamically moving test masses\cite{Cooper2022,Carter2025}, an effect also seen for other sensors~\cite{Hines2023}.
\par
Our goal is to assess if the same noise floor as that of a rigid piece can be measured for a moving mirror, and validate that the motion tracked by the sensor matches the actual motion of the moving test mass. To do this we need a mirror whose motion we can control and know well. A standard method of controlling length in optical setups is a piezoelectric stack mounted behind a mirror, however, such stacks are prone to substantial nonlinearities themselves. This could be mitigated by probing both sides of the mirror and measuring the common motion. However, if two HoQIs were used for such a measurement, we would not be able to assess their common behaviour, while if a HoQI and a different sensor were used, we would not be able to distinguish between the HoQI's nonlinearities and the other sensor's.
\par
\begin{figure}
    \centering
    \includegraphics[width=1\linewidth]{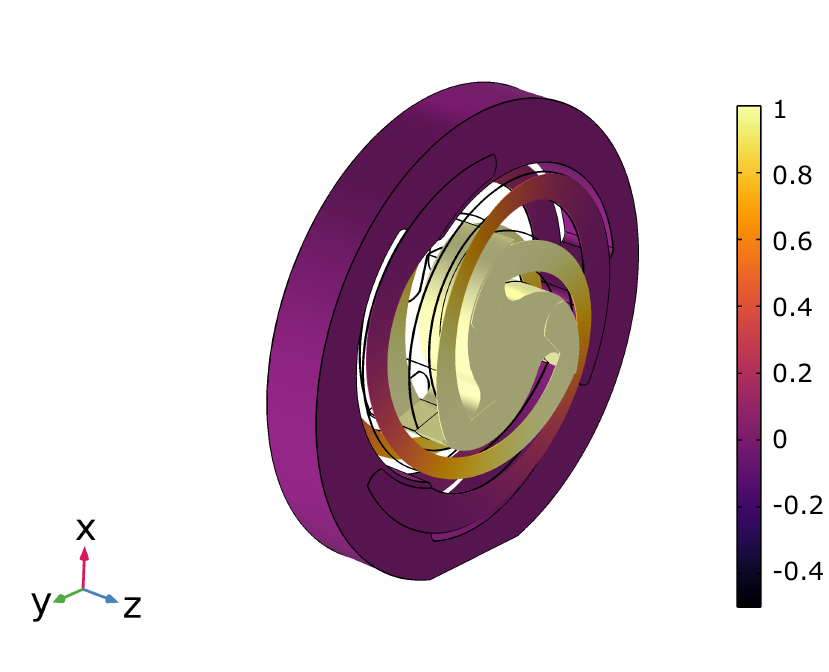}
    \caption{Mode shape of the mechanical resonator used to induce arm length changes in the HoQI. The axis indicates the local displacement as a fraction of maximum motion. The resonator was set up such that the z-axis was aligned to the incident beam from the HoQI, such that motion in this direction was measured. The resonator is a monolithic fused silica piece, realised using direct bonding, to achieve the high Q factor. Its test mass weighs 3\,g. A gold coating has been applied to the centre of the piece to reflect incident laser light.}
    \label{fig:resmode}
\end{figure}

Instead, we use a fused silica mechanical resonator, whose mechanical motion is well defined to be linear for the displacement amplitudes investigated here.  The resonator is described in Ref.~\cite{Birckigt2024}. The resonator's fundamental mechanical mode is heavily constrained to motion in one cartesian direction. The mode shape and resonator design are shown in Fig.~\ref{fig:resmode}. The fundamental mode of the resonator is at 20.03\,Hz, whilst the next lowest mode is a tip tilt mode at 300\,Hz, well outside our measurement band. In our experiments, we use the resonator as the test mass of the HoQI. The resonator is set up so that the coordinate system from Fig.~\ref{fig:resmode} matches that of Fig.~\ref{fig:hoqilayout}, meaning its $z$ motion is the measured motion of the HoQI. The whole setup is placed on an optical table which is seismically isolated and controllable~\cite{Wanner2012}.\par
By applying an impulse or ringing up the table, we can excite the resonator's mode to different amplitudes. Due to the resonator's high mechanical quality factor (150,000), it will then hold a similar amplitude for an extended time after the input signal is turned off, so we can do studies at the amplitude of motion without external disturbances. Using such a resonator gives us a well-defined, controllable arm length change. 
\subsection{Measuring Nonlinearities in Displacement Sensors}
\label{subsec:measuring_nonlinearities}

For displacement sensors such as HoQIs that rely on measuring two quadratures to recover the true phase, in the ideal case the data draws a sector of a circle, and the phase can be recovered simply by using the arctangent function, as detailed in Sec.~\ref{subsec:reconstructing phase signals}.

However, in realistic cases there are imperfections leading to errors in the phase measurement on individual PDs. This can arise, for instance, from imperfections in the manufacturing or alignment of the polarisation optics. The relevant optics can be beam splitters, half-wave plates or quarter-wave plates, with the latter two the most significant culprits according to investigation in Ref.~\cite{Cooper2020}. Transient sources of nonlinearities are also possible, induced by thermal effects or physical motion of the device affecting alignment.

The result of these imperfections is that, in general, the data draws a Lissajous figure. In the typical case, where disturbances change slowly compared to the oscillation frequency, the data draws an ellipse. A general ellipse can be described with five parameters: two offsets from the center $C_1$ and $C_2$ in $Q_1$ and $Q_2$, respectively, a rotation angle $\theta$ with respect to the $Q_1$ axis, and two radii $R_1$ and $R_2$ along $Q_1$ and $Q_2$, respectively. However, considering four parameters is sufficient, because only the radius ratio $R_1/R_2$ affects the angle $\phi$ on the quadrature plane, while the absolute value of the radius is irrelevant for the phase readout. 

Directly using Eq.~\ref{eq:arctan} without any prior correction of the data will introduce a non-linear error in the phase and its dependent, the distance measurement. In our experimental setup, we could directly observe the effect of the nonlinear error. There happened to be a mechanical resonance at a frequency of around 20.7\,Hz, originating in the mounting or in the seismic isolation system. Due to this, a non-linear response of the HoQI to a change in the optical phase introduces noise at a beat note at the difference in frequency between this mode and our resonators 20.03\,Hz mode. This creates a small peak at 0.7\,Hz, in addition to harmonics of the 20.03\,Hz peak. Therefore, the height of the 0.7\,Hz peak and other harmonics in the amplitude spectral density act as a gauge of the nonlinearity-induced optical phase error. In the ideal case with no error, these peaks should nearly vanish. To further expose this effect, we subtracted the signals from additional witness sensors measuring motion across our experimental setup. Those sensors were six L-4C seismometers measuring the motion of the isolated platform and one STS-2 seismometer measuring ground motion. The coherent subtraction used a Multi-Channel Coherent Subtraction (MCCS) routine as described in~\cite{Allen1999,spicypy}. The prominence of the 0.7\,Hz peak in the amplitude spectral density (ASD) after coherent subtraction is used as a figure of merit for non-linear errors throughout this paper. Additionally, to quantitatively assess residual non-linearity, we calculate the noise level in ASD after coherently subtracting witness sensors' signals at around 40.06\,Hz, the second harmonic of the 20,03\,Hz peak.

\section{Realtime Compensation for Nonlinearities in HoQI Readout}
\label{sec:compensating_nonlinearities}
To use HoQIs for seismic isolation control in a gravitational wave detector \cite{Matichard2015a,Matichard2015b}, we need a means of real-time suppression of nonlinear behaviour. Therefore, we need a simple algorithm that can be applied in sensor readout digitally in real-time to correct the nonlinear behaviour of the HoQI. In this section, we present such an algorithm. Several parameters are needed in this algorithm that must be estimated. We present how to estimate these parameters using data from the HoQI readout.
\subsection{An Algorithm for Correcting Nonlinearity in HoQIs}
During quiet conditions achieved in our system, test mass motion at the resonance frequency does not exceed one full FSR. Therefore, to find the initial ellipse parameters and, in parallel, to measure the Q-factor of the oscillator, a ringdown measurement was performed. The motion at resonance was enhanced by exciting the horizontal actuators of the isolation platform at 20.03\,Hz until the HoQI measured a periodic motion of 5 FSRs. Then the excitation was turned off and the oscillation was left to decay due to friction.

The ellipticity was initially corrected by the following algorithm:
\begin{itemize}

\item Remove offsets in quadradures, $Q_1^{'} = Q_1 - C_1$, $Q_2^{'} = Q_2 - C_2$
\item Apply scaling factors to quadratures, $Q_1^{''} = S_1 \cdot Q_1^{'}, Q_2^{''} = S_2 \cdot Q_2^{'}$
\item Correct for phase error in one quadrature, $\alpha$, by applying the following transform: $Q_2^{'''} = \frac{Q_2^{''} + Q_1^{''}\cdot \text{sin}(\alpha)}{\text{cos}(\alpha)}$.
\end{itemize}

The last step in this transform is a consequence  of correcting the phase error in $Q_2^{''}$ using the trigonometric identity: $\text{cos}(\phi+\alpha) = \text{cos}(\phi)\text{cos}(\alpha) - \text{sin}(\phi)\text{sin}(\alpha)$. These operations are implemented in our digital data acquisition system (Control and Data System, CDS) and used for real-time correction of the distance signal. The parameters are not the same as in usual ellipse fitting, but were convenient to be estimated manually at the time. Since both options are mathematically equivalent, only usual ellipse parameters will be considered in the following.
\begin{table}

    \centering
    \small
\begin{tabular}{  l | l | l | l | l }
 \multicolumn{5}{l}{Ellipse parameters from ringdown:} \\
  $C_1,\ V$ & $C_2,\ V$ & $R_1,\ V$ & $R_2,\ V$ & $\theta,\ \text{rad}$  \\ 
  $0.2204$ & $0.1529$ & $7.83681$ & $8.50401$ & $0.17122$  \\
 \multicolumn{5}{l}{Ellipse parameters from the fit:} \\
  $C_1,\ V$ & $C_2,\ V$ & $R_1,\ V$ & $R_2,\ V$ & $\theta,\ \text{rad}$  \\
     \makecell[l]{0.20569 \\ $\pm {10}^{-5}$} & \makecell[l]{0.14190 \\ $\pm {10}^{-5}$} & \makecell[l]{7.85903 \\ $\pm {10}^{-5}$} & \makecell[l]{8.51400 \\ $\pm {10}^{-5} $} & \makecell[l]{0.15962 \\ $\pm 2\cdot {10}^{-5}$}  \\
\end{tabular}
\normalsize
\caption{Parameters for correction of quadratures $Q_1, Q_2$. Parameters found manually from ringdown measurement were fixed throughout the measurements presented in this paper. Ellipse parameters from the fit shown here were found on roughly one minute of near one FSR motion data.}
\label{tab:ellipse_parameters}
\end{table}
Assuming the ellipse traced is static, one can calculate these values once and then have real-time correction for all future measurements. Which method and which accuracy is required for the parameter estimation, depends on the respective position and length of the ellipse section. We present several methods for estimating these values in the following sections. 
\subsection{Simple Ellipse Fitting}
To validate the initial parameter estimation and live-correction method, we excited the isolation platform again to create one hour of motion data with a near-constant amplitude of just under one FSR. This measurement was taken over one month after the initial ringdown measurement. Then ellipse fitting was attempted in post-processing.

\begin{figure}
    \centering
    \includegraphics[width=0.43\textwidth]{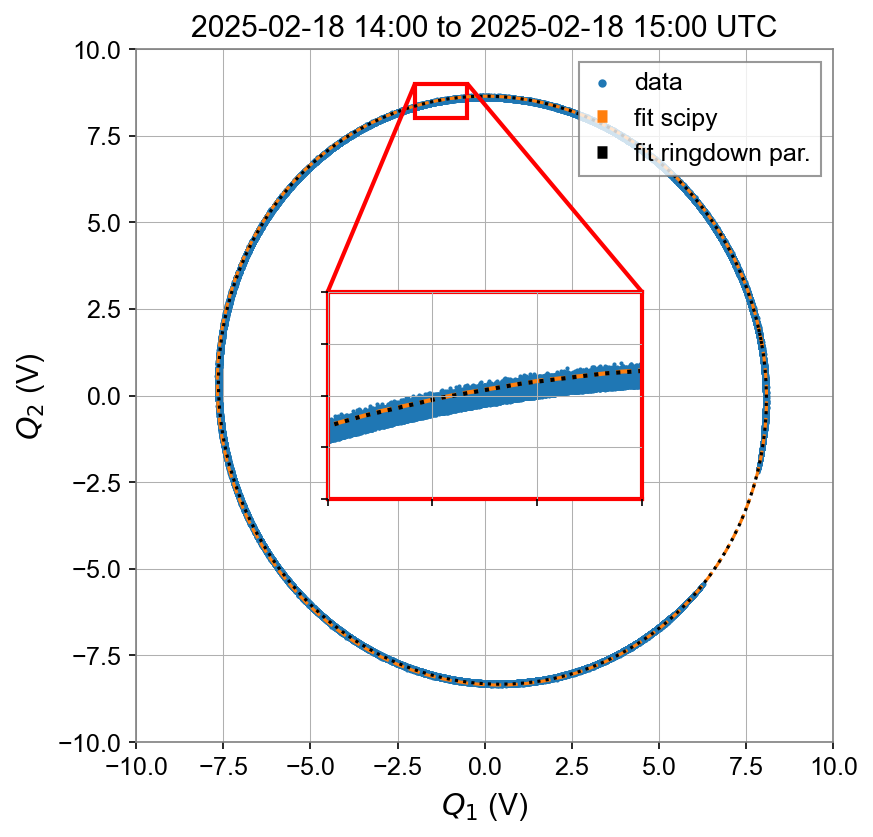}
    \caption{Trace drawn by the data on $Q_1 Q_2$-plane during an hour of near one FSR motion (blue scatter), ellipse fit performed with parameters from ringdown measurement (``ringdown par.'', black dotted line), and ellipse fit performed with \texttt{scipy.curvefit} routine (``fit \texttt{scipy}'', orange dashed line) on only one minute of these data in post-processing.}
    \label{fig:ellipse_high_motion}
\end{figure}

\begin{figure}
    \centering
    \includegraphics[width=.45\textwidth]{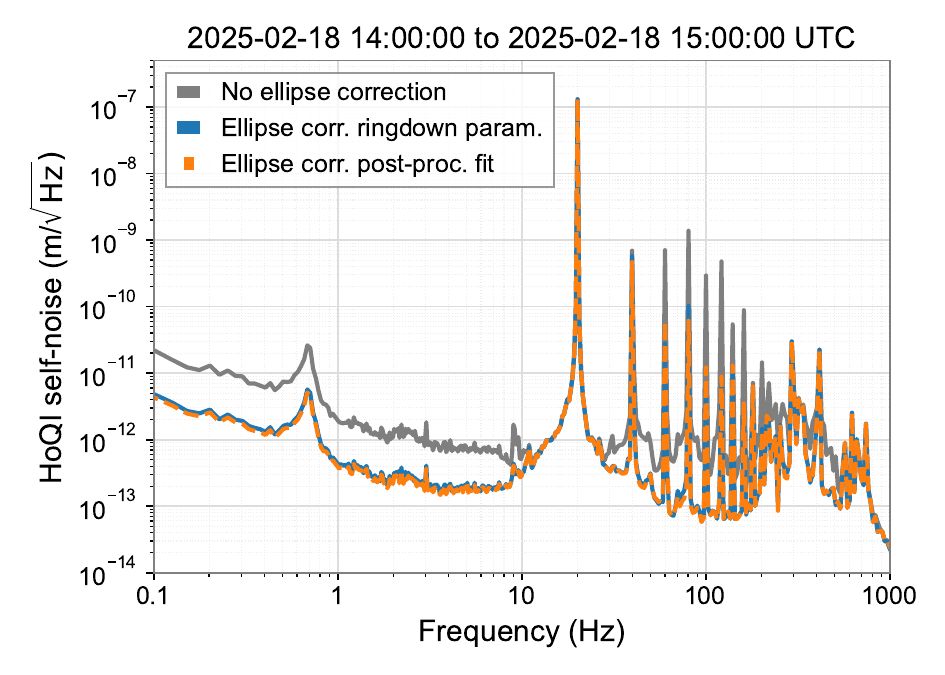}
    \caption{ASD of HoQI self-noise (after coherently subtracting motion registered by the witness sensors) for one hour of near one FSR motion data. Ellipse correction with parameters from ringdown measurement (blue line) is compared against a fit with \texttt{scipy.curve\_fit} on roughly the first minute of data, with correction applied then to the full hour of data (orange line) and no ellipse correction (gray line).}
    \label{fig:asd_high_motion}
\end{figure}

For this amplitude of motion, fitting the ellipse directly using points of data on the $Q_1 Q_2$-plane to an idealised circle works well, taking only seconds of computation on a typical desktop, using \texttt{scipy.curve\_fit} algorithm. This algorithm takes a generic user-defined fitting function, in our case the equation describing a general ellipse.
We have also tried an algorithm for ellipse fitting based on the least-squares method~\cite{halir1998} and implemented in python in~\cite{hammel2020}. That algorithm produced the same ellipse as \texttt{scipy.curve\_fit} and was even faster. It is also simple enough that it can be implemented in the future directly in CDS as a program in C for automated calibration.
\par
Data points for one hour of data on the $Q_1 Q_2$-plane are shown in Fig.~\ref{fig:ellipse_high_motion}, along with the ellipse fit with ringdown parameters and the ellipse fit in post-processing. In Fig.~\ref{fig:asd_high_motion} an amplitude spectral density with coherently subtracted witness sensors is shown, comparing raw data to a correction with ellipse parameters found in the ringdown measurement and to an ellipse fit in post-processing, specified in Tab.~\ref{tab:ellipse_parameters}.

It can be seen that the noise floor is significantly reduced after ellipse correction. It can also be seen that parameters from the ringdown measurement are already close to the optimal fit for this data, even though they were obtained during different data taking period. This suggests that ellipse parameters are stable to the first order, and therefore constant correction parameters can be used for distance measurement, as we have done in Ref.~\cite{Carter2025}. 

\subsection{Whitening filter for nanometre-scale motion}
\begin{figure}
    \centering
    \includegraphics[width=.48\textwidth]{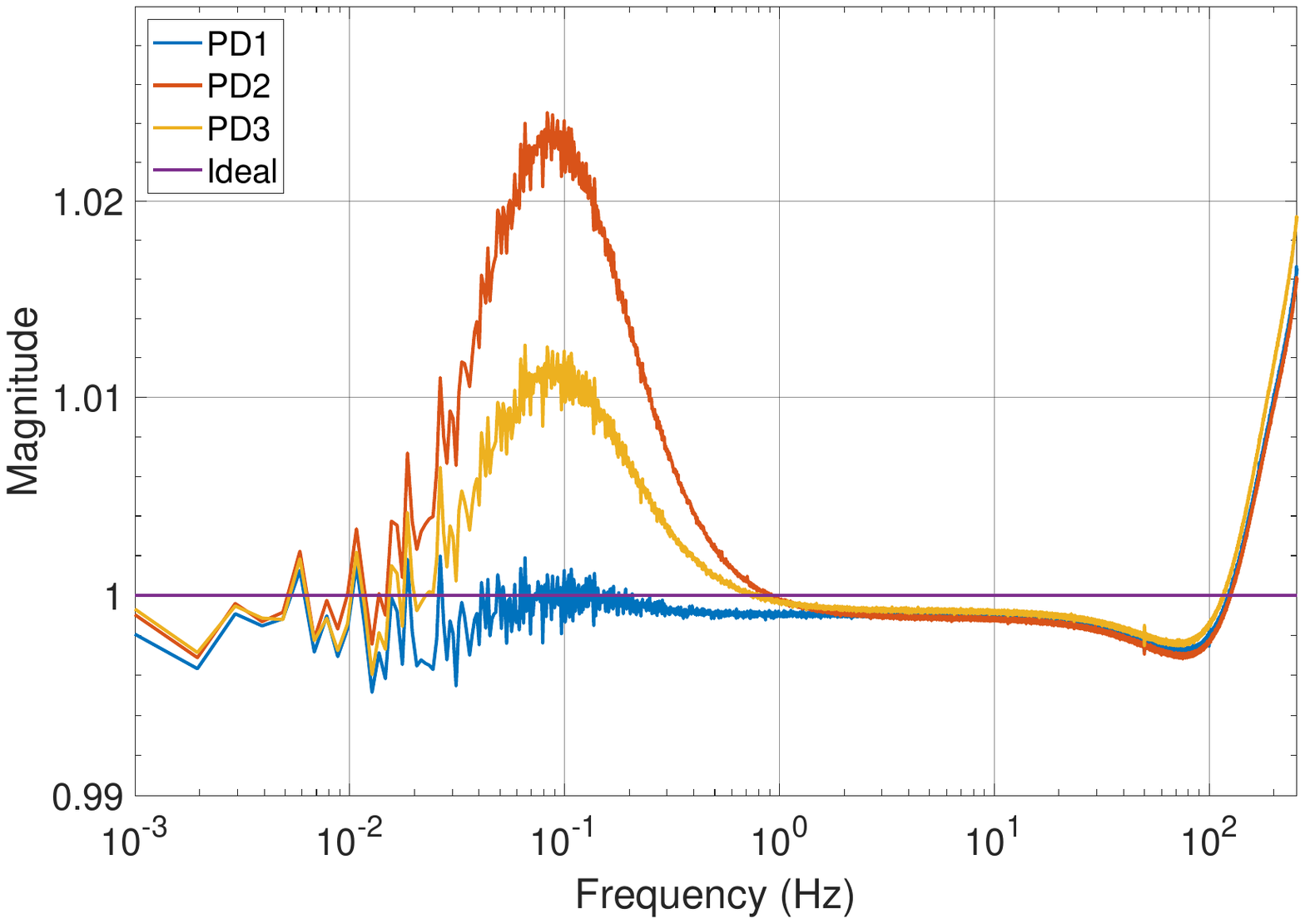}
    \caption{Measurement of the transfer function of the whitening filters for each PD channel multiplied with the previously calculated digital de-whitening filter.}
    \label{fig:whitetf}
\end{figure}

\begin{figure}
    \centering
    \includegraphics[width=.48\textwidth]{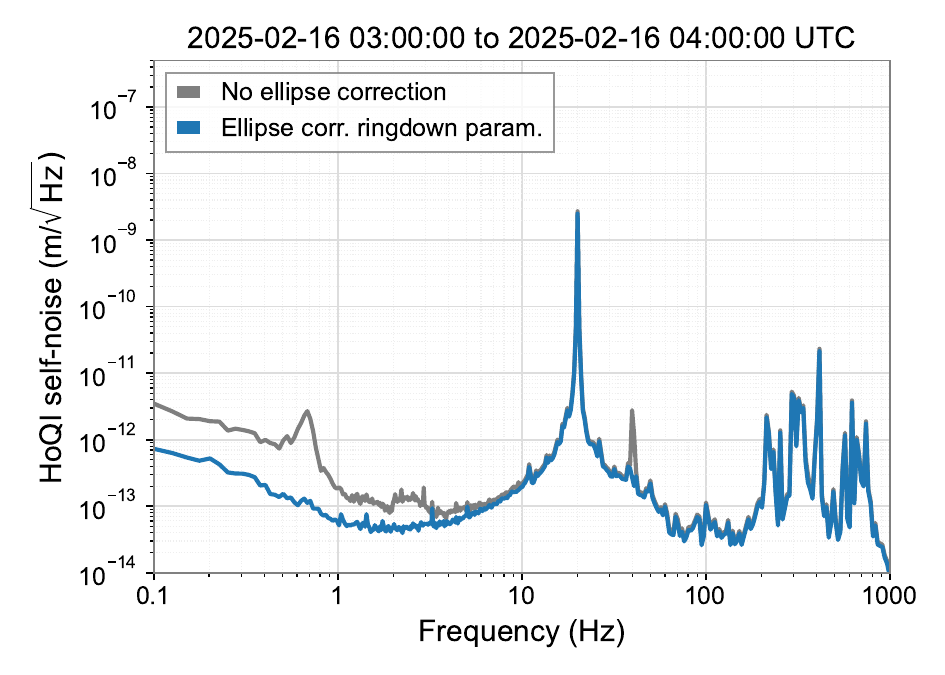}
    \caption{ASD of HoQI self-noise, after coherently subtracting motion registered by the witness sensors, for one hour of nanometre-scale motion data. Ellipse correction with parameters from ringdown measurement (blue line) is compared to no ellipse correction (gray line). Analogue whitening and subsequently digital de-whitening filters have been used for this measurement.}
    \label{fig:asd_tiny_motion}
\end{figure}

During one measurement with especially stable state we had motion on resonance always less than 5\,nm, and often even below 1\,nm. Due to the very small motion amplitude, the Analogue to Digital Converters (ADCs) were not used efficiently with respect to their limited dynamic range. When using ADCs, it is common to amplify the signal at frequencies where it is low before it is digitised. The filter stages are called whitening filters because they make the signal amplitude less frequency dependent. This way, the dynamic range of the ADC is used more efficiently. After digitisation, the whitening filter is inverted by a digital de-whitening filter to recover the original signal. While it is usually not critical to match the two filters perfectly, in the case of HoQIs, any error in the de-whitening filter can lead to amplitude and phase mismatch between different PD signals. Those errors introduce new nonlinearities in the displacement readout of the HoQI. 
For the first test of a whitening filter in combination with a HoQI, we chose corner frequencies which are as far away from the resonance frequency of the oscillator as possible, so that potential errors in the signals were easier to understand. This resulted in a band pass filter with a calculated gain of 3.2 between the pole frequencies of 0.159\,Hz and 723.4\,Hz. 
With a gain of 3.2 at the resonance frequency of 20.03\,Hz, even the previously shown sub-FSR motion could lead to saturations in the amplifier. To prevent this, the isolation platform was set to a control state which actively suppressed its motion in a narrow band around 20.03\,Hz. This reduced the maximum amplitude of the oscillator motion at its resonance to 5\,nm.
Despite the conservative whitening filter design, their usage initially led to a significant increase of nonlinearities in the HoQI readout when the de-whitening filters were designed purely based on the calculated filter response. To investigate this, output channels of the CDS were connected to the inputs of the amplifier instead of the PDs, so that transfer functions of each channel could be measured. The results are shown in Fig.~\ref{fig:whitetf}.
Below 1\,Hz, the channels have different responses due to tolerances in electrical components shifting the pole and zero frequencies of the whitening filters. This was compensated by correcting the de-whitening filters based on this measurement. The artefacts above 100\,Hz are related to the sampling rate available for this measurement; they are also present without whitening filters, so they were not compensated.
As a result, the nonlinearities of the HoQI response could be reduced to a similar level as without using whitening filters. The comparison was done without changing any ellipse parameters. The measurements presented in Ref.~\cite{Carter2025} were taken in this state, so that the HoQI sensitivity was improved by a factor of 3 in the band of the whitening filter. The noise floor in this state, before and after ellipse correction, is shown in Fig.~\ref{fig:asd_tiny_motion}.

\section{Residual non-linearity correction in post-processing}\label{sec:postcorrect}
The previous section presented methods of what could be suppressed in a real-time system used, for example, in active control. In systems that measure displacement signals that are then reconstructed later \cite{Kang2006, Heijningen2023}, we can correct even more errors in post-processing, as we can effectively correct the ellipse for temporal changes and hysteretic effects. Methods of further nonlinearity suppression in post processing are detailed in this section.
\subsection{Hysteresis compensation during near full FSR motion}
\label{subsec:near_full_fringe_motion}

During  motion with an amplitude near one full-FSR, once the data are corrected with a fitted ellipse, the result is very close to an ideal circle on the $Q_1 Q_2$-plane. To assess how close exactly, we calculated radius deviation, $R_d$, as a function of phase by subtracting the mean radius for each phase:
\begin{equation}
\label{eq:radius_deviation}
\begin{aligned}
& \phi = \text{arctan2}(Q_1,Q_2),\\
& R(\phi) = \sqrt{Q_1^2+Q_2^2},\\
& R_d(\phi) = R(\phi) - \overline{R}.
\end{aligned}
\end{equation}

\begin{figure}
    \centering
    \includegraphics[width=.48\textwidth]{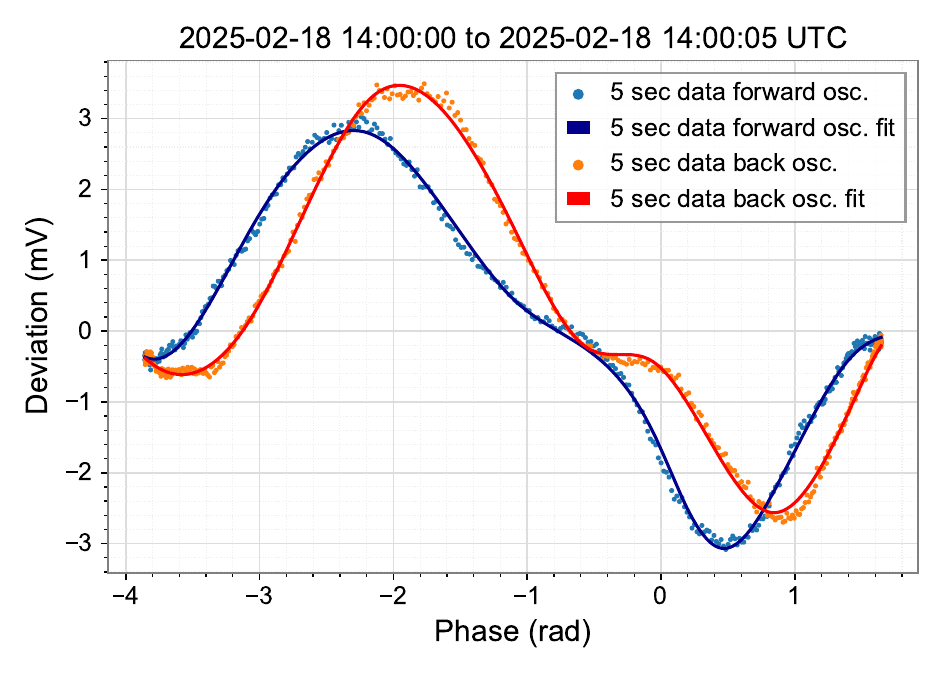}
    \caption{Residual non-linearities in data shown as a deviation from a perfect circle.}
    \label{fig:hysteresis}
\end{figure}

The result for one oscillation is shown in Fig.~\ref{fig:hysteresis}. The direction of each half-oscillation is shown with different color, and it can be clearly seen that the data traces different paths depending on direction. This effect has been confirmed for every oscillation. A possible reason could be a cross coupling from a translation of the test mass to a small tilting motion, for example due to thickness deviations in the flexures holding the test mass. Such a tilting motion could change the signal on the PDs due to a change in fringe visibility of the interferometer or due to an inhomogeneous responsivity of the PD across its surface.
Therefore, any further correction has to take directionality into account.

First, we tried fitting data to a different ellipse depending on the direction of motion.
For that, the data was split into two data sets depending on the direction, and for each data set, an ellipse fit was performed. Subsequently, the corrected phase time series were stitched together. To avoid introducing noise from the thereby created step functions, the phase time series were filtered and downsampled before further processing. Nevertheless, the result was worse than fitting all data with a single ellipse. The ``nonlinearity gauge'' at 0.7\,Hz and 20.03\,Hz harmonics did not get reduced; moreover, there was additional broad-frequency noise injected, possibly because of the stitching procedure. The reason why the nonlinearity peaks did not get reduced is possibly due to a deviation from an ellipse in the vicinity of turning points in the hysteresis curve, where two trajectories must connect, which is dependent on motion amplitude and hence is always at a slightly different phase.

Subsequently, we used ellipse correction with parameters from the ringdown measurement, and have fitted the residual deviations for  each motion direction separately with splines, also shown in Fig~\ref{fig:hysteresis}, using \texttt{make\_splrep} routine from \texttt{scipy} package. To avoid overfitting, a smoothing condition, $s$ was chosen such that the fit captures the general trend of the data but not the noise by following the points too closely. The condition $s=3\cdot 10^{-6}$ was found to give the best results.

Using this fit of deviations, $R_d^S(\phi)$, the phase has been corrected:
\begin{equation}
\label{eq:phase_correction}
\begin{aligned}
& Q_1^{'} = \overline{R}\cdot \text{sin}\left(\phi \cdot C \cdot (1 - R_d^S(\phi)\right), \\
& Q_2^{'} = \overline{R}\cdot \text{cos}\left(\phi \cdot C \cdot (1 - R_d^S(\phi)\right), \\
& \phi^{'} = \text{arctan2}(Q_1^{'},Q_2^{'}).\\
\end{aligned}
\end{equation}

The calibration constant, $C$, was needed because the function $R_d^S(\phi)$ has unit of volts, while the phase has unit of radians. This constant was found empirically by choosing the value that gives the best reduction of prominence around 0.7\,Hz peak. It was found to be $C=2.615\ \text{rad}/\text{V}$.

\begin{figure}
    \centering
    \includegraphics[width=.48\textwidth]{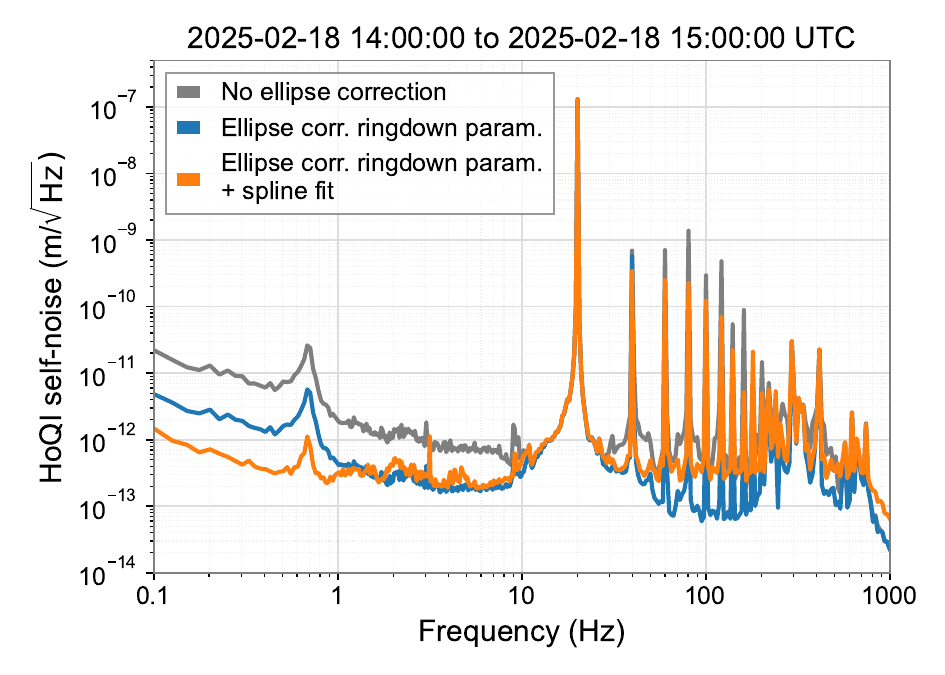}
    \caption{ASD of HoQI self-noise (after coherently subtracting motion registered by the witness sensors) for one hour of sub-FSR motion data. Ellipse correction with parameters from ringdown measurement (blue line) is compared to hysteresis correction with spline fit (orange line).}
    \label{fig:asd_hysteresis}
\end{figure}

The correction in Eq.~\ref{eq:phase_correction} was done separately for each direction of motion, using the corresponding fit $R_d^S(\phi)$. 
The result was a suppression of the 0.7\,Hz peak and the harmonics of the 20.03\,Hz peak, especially visible for 40.06\,Hz, shown in Fig~\ref{fig:asd_hysteresis}. Unfortunately this correction introduced its own error, manifesting as a higher noise floor at high frequencies due to imperfect fitting and perhaps residual effects of step functions in the time series. Therefore we have not applied it to all data, but it serves as a proof-of-principle that nonlinearity due to hysteresis-like behavior can also be corrected.

\subsection{Repeated ellipse fitting during sub-FSR motion}
\label{subsec:sub_FSR_motion}
\begin{figure*}
    \centering
    \subfigure[First time segment]{\includegraphics[width=0.43\textwidth]{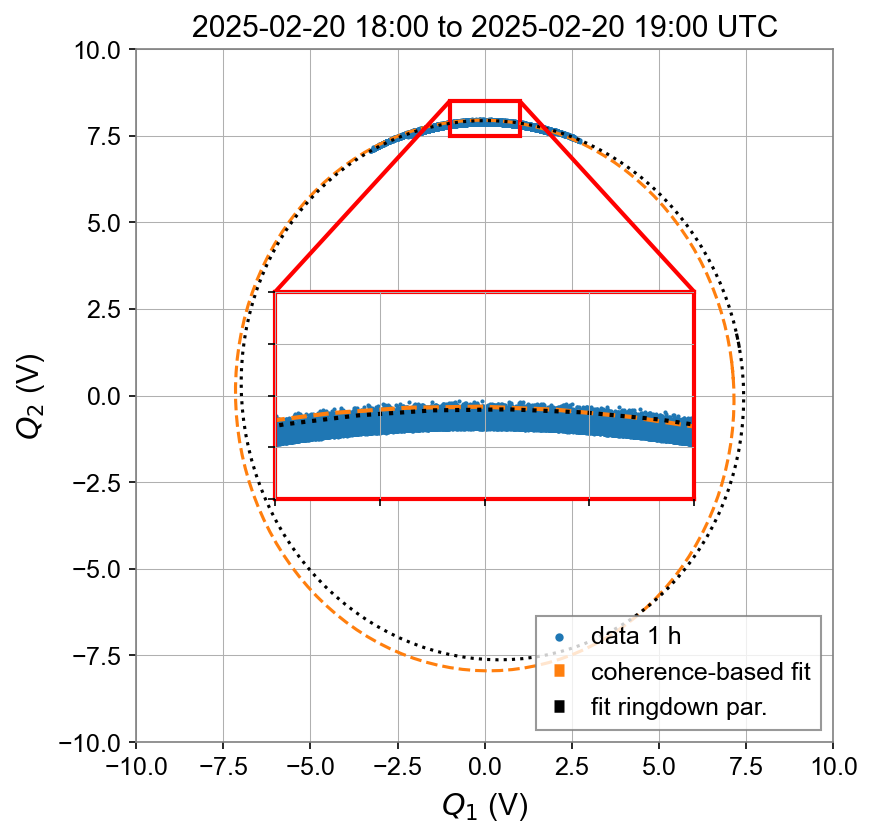}
    \label{fig:ellipse_intermediate_motion1}
    \hfill}
    \subfigure[Second time segment]{\includegraphics[width=.43\textwidth]{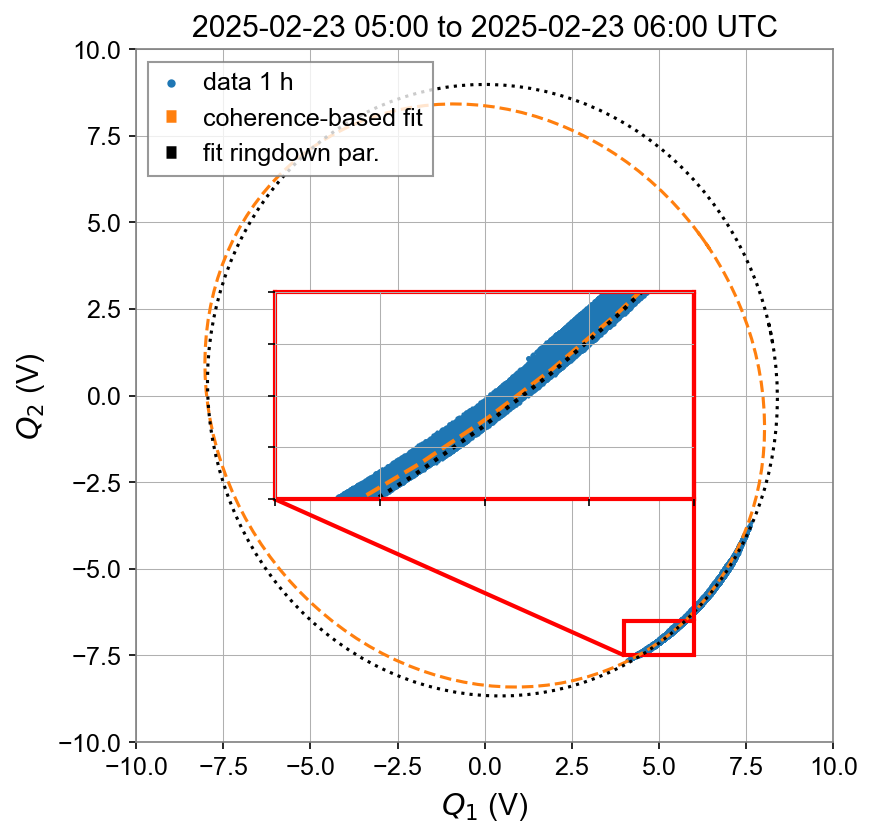}
    \label{fig:ellipse_intermediate_motion2}}
    \caption{Traces drawn by the data on $Q_1 Q_2$-plane during an hour of sub-FSR motion (blue scatter), ellipse fit performed with parameters from ringdown measurement (``ringdown par.'', black dotted line), and ellipse fit performed with nonlinear peak prominence-based fit (``coherence-based fit'', orange dashed line) in post-processing.}
\end{figure*}

Sub-FSR motion data were collected in a time span of roughly 3.5 days over the weekend. The typical motion amplitude over one hour of data is shown in Fig.~\ref{fig:ellipse_intermediate_motion1} including the ellipse fit. On one hour time scale, the motion amplitude is dominated by the 20.03\,Hz oscillation, but over the course of the full measurement, the data position slowly rotated on the $Q_1 Q_2$-plane, eventually doing close to one full rotation. As another example, another hour of data is shown in Fig.~\ref{fig:ellipse_intermediate_motion2}, nearly on the opposite part of the phase ellipse.

We have found that there is a non-negligible variation of ellipse parameters on a time scale of hours. However, unlike near full FSR motion data, fitting here could not be done using only data on the $Q_1 Q_2$-plane, as fitting errors exceeded the typical ellipse change. In other words, for motion of much less than one FSR, there are multiple ellipses that can be drawn through the same data with nearly equal mean square errors, but with significantly different parameters. In particular, the radius ratio and center offsets become poorly constrained.

\begin{figure*}
\centering
    \subfigure[Ellipse correction with parameters from ringdown]{\includegraphics[width=0.48\textwidth]{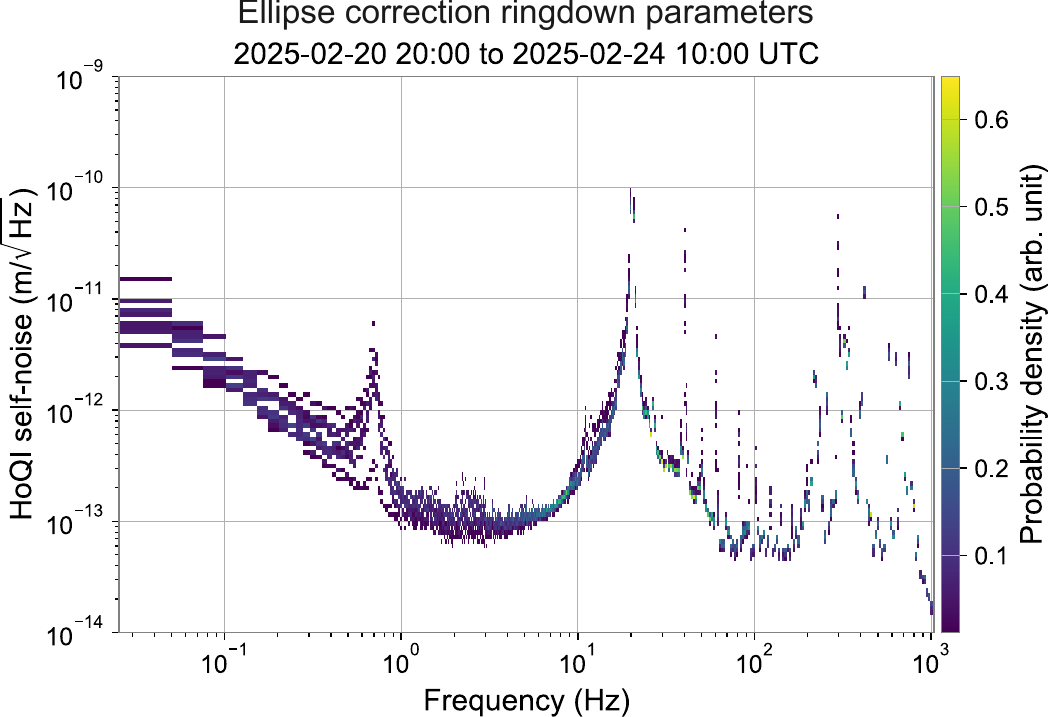}
    \label{fig:pasd_live}
    \hfill}
    \subfigure[Ellipse correction with parameters from fit to each hour]{\includegraphics[width=.48\textwidth]{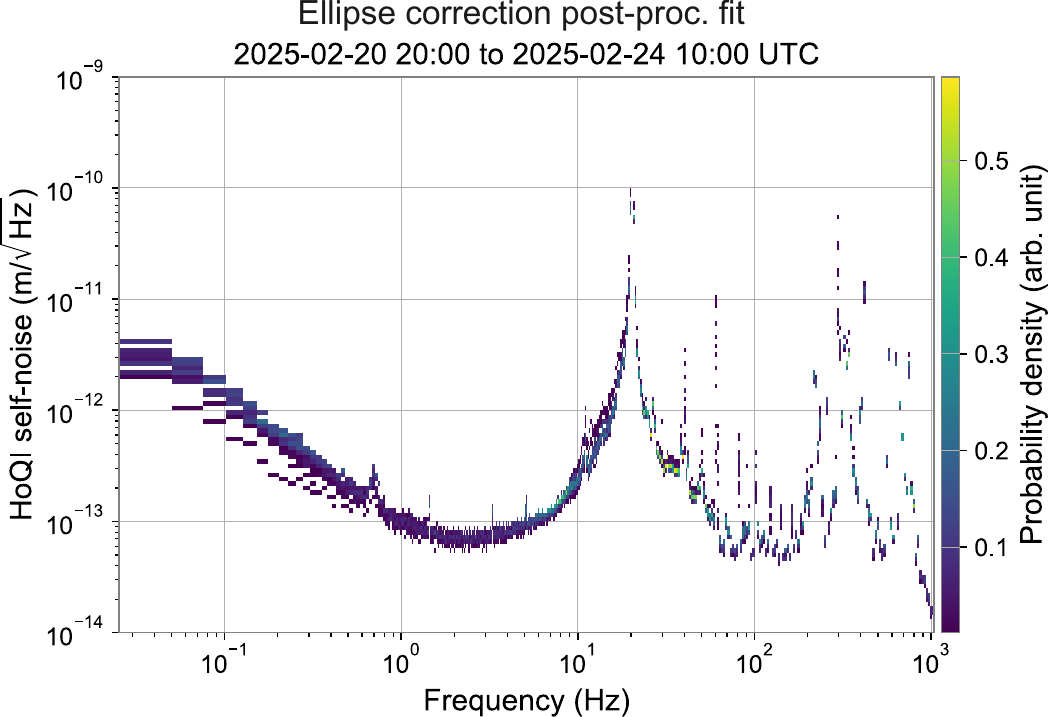}
    \label{fig:pasd_fit}}
    \caption{Probabilistic ASD of HoQI self-noise, each ASD is one-hour duration with the color and spread showing more common (probable) amplitude values.}
\end{figure*}

Subsequently instead of the $Q_1 Q_2$-plane we have used a fit based on the result of coherent subtraction for one hour of data. We have used a loss function based on the prominence of the 0.7 Hz ``nonlinearity gauge'' peak, defined as:

\begin{equation}
\begin{aligned}
\label{eq:fit_loss}
& L(Q_1, Q_2, R_1/R_2, \theta) = \\
& = \frac{1}{N(Q_1, Q_2, R_1/R_2, \theta)} \int_{f_1}^{f_2} S_{xx}(Q_1, Q_2, R_1/R_2, \theta) \,df,
\end{aligned}
\end{equation}

where $S_{xx}$ is the residual PSD of the HoQI distance estimated with ellipse correction using given $R_1/R_2$ and $\theta$ parameters and coherently subtracted witness sensor signals, $f_1 = 0.3\ \text{Hz}$ and $f_2 = 1.0\ \text{Hz}$ are frequencies that serve as boundaries of the 0.7 Hz peak, and the normalizing coefficient $N$ is defined as follows:

\begin{equation}
\begin{aligned}
& N(Q_1, Q_2, R_1/R_2, \theta) = \\ 
& =\ \int_{0}^{f_1} S_{xx}(Q_1, Q_2, R_1/R_2, \theta) \,df + \\
& + \ \int_{f_2}^{f_n} S_{xx}(Q_1, Q_2, R_1/R_2, \theta) \,df,
\end{aligned}
\end{equation}

where  $f_n = 1024\ \text{Hz}$ is the Nyquist frequency for the sampling rate of $2048\ \text{Hz}$ at which the fitting was performed, down-sampled from $16384\ \text{Hz}$ sampling frequency at which individual PD signals were recorded. The integration covers ranges outside of the peak, and the division by $N$ 
in Eq.~\ref{eq:fit_loss} quantifies peak prominence. 
The residual PSD $S_{xx}$ is calculated with quadrature values scaled and rotated by the corresponding fit parameters:

\begin{equation}
\begin{cases}
    Q_1^{'} = (R_1/R_2) \cdot \left(Q_1 \text{cos}(\theta) + Q_2 \text{sin}(\theta) \right) \\
    Q_2^{'} = -Q_1 \text{sin}(\theta) + Q_2 \text{cos}(\theta).
\end{cases}
\end{equation}

However this by itself was not yet sufficient for a robust fit. We have found that four ellipse parameters produced a too large phase space for the coherent-subtraction-based fit. This could result in a non-physical ellipse, e.g. instead of a nearly-circular ellipse, fitting data by a near straight-line ellipse with extreme radius ratio. It could also lead to non-optimal solutions that only increased noise. The introduction of tight parameter constraints could somewhat improve the results but it is discouraged as we did not know \textit{a priori} what could be the maximum ellipse parameter variation. Besides, we found that the fit would frequently pick the values of parameters right at the constraint border, suggesting that it essentially failed to find the optimal solution within the constraints.

\begin{figure}
    \centering
    \includegraphics[width=.48\textwidth]{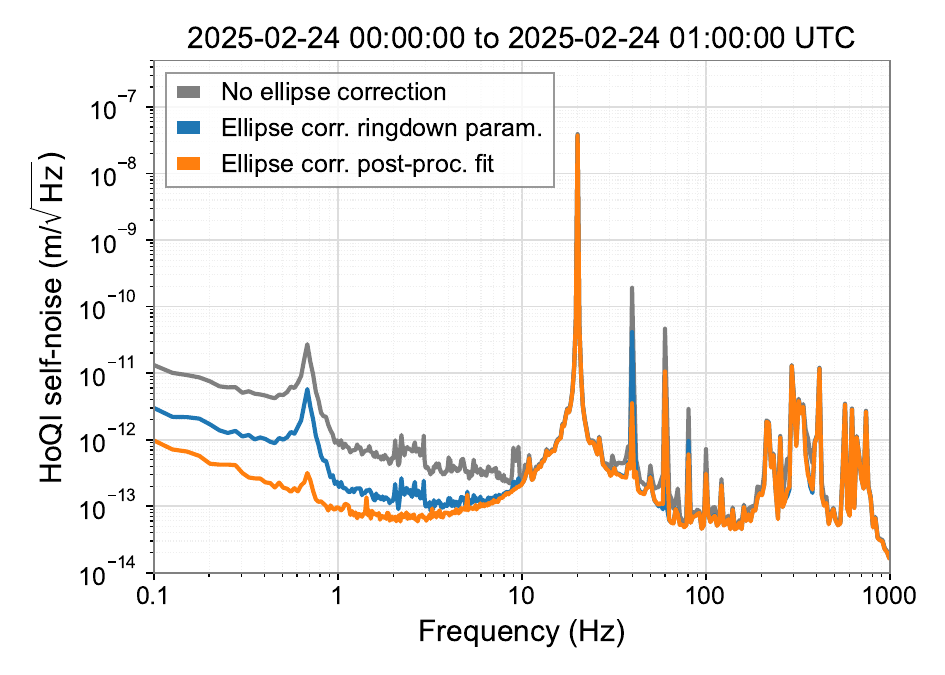}
    \caption{ASD of HoQI self-noise, after coherently subtracting motion registered by the witness sensors, for one hour of sub-FSR motion data. Ellipse correction with parameters from ringdown measurement (blue line) is compared against a coherent-subtraction-based fit on that hour of data (orange line) and no ellipse correction (gray line).}
    \label{fig:asd_intermediate_motion}
\end{figure}

To counteract these effects, we have used a fit where only two parameters are varied: radius ratio $R_1/R_2$, and rotation angle $\theta$. We have also used \texttt{scipy.optimize.brute} algorithm, sampling an initial grid of $100 \times 100$ points, for the first fit spanning $0.95 < R_1/R_2 < 1.05$, $0.0\ \text{rad} < \theta < 0.1\ \text{rad}$, and for subsequent fits centered on the previous fit optimal point with the same range. After sampling this initial grid, the fit continued to find the optimal solution using Nelder-Mead simplex algorithm. Solutions outside of the grid were also allowed, if they are the result of this algorithm. Criteria to determine convergence were set as $10^{-4}$ for the maximum change in both loss function values and the arguments $R_1/R_2$ and $\theta$. We have found that such a setup gives a robust result without unphysical solutions. 

We have used this fitting for each hour of data over roughly 3.5 days. Each fit took around 10 minutes on Intel Core i7-8700 Processor (6 cores, 3.20 GHz), using multiprocessing to sample the initial grid. The initial grid sampling was faster than the subsequent fitting, taking only a fraction of the 10 minutes. 

The results are shown in Fig.~\ref{fig:pasd_fit} as a probabilistic amplitude spectral density of coherently-subtracted witness sensors from the HoQI signal and can be directly compared to Fig.~\ref{fig:pasd_live} where parameters from the ringdown measurement were used. It can be clearly seen that the noise floor at low frequencies is reduced, with much less spread. The 0.7\,Hz and 40.06\,Hz peaks are strongly suppressed, indicating highly reduced non-linear noise coupling. One example where the coherent-subtraction-based fit clearly surpasses performance of the ellipse correction with parameters from the ringdown measurement is shown in Fig.~\ref{fig:asd_intermediate_motion}.

\begin{figure}
    \centering
    \includegraphics[width=0.48\textwidth]{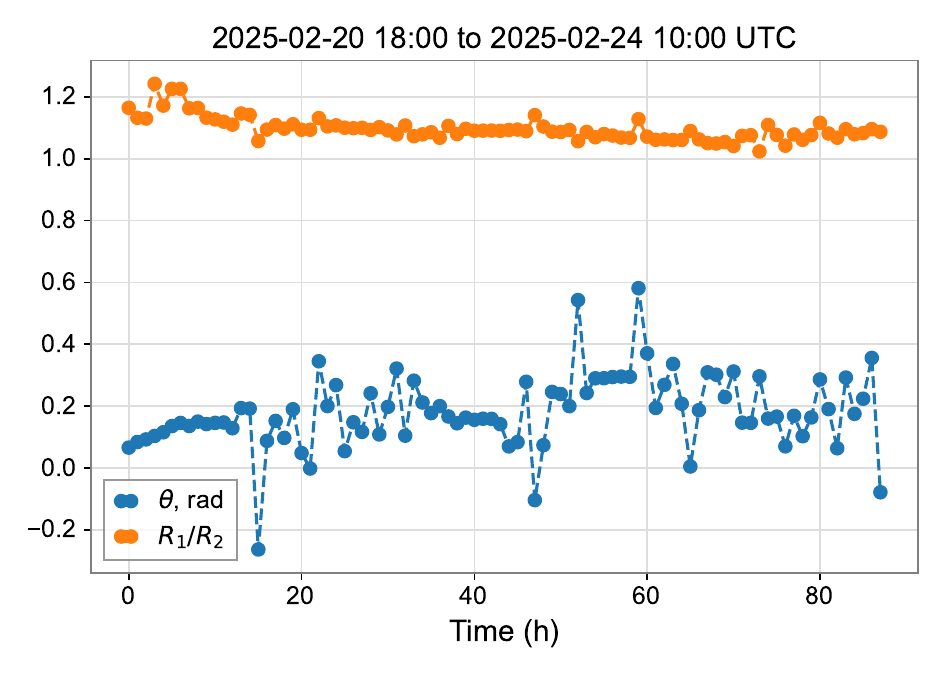}
    \caption{Variation of ellipse parameters over roughly 3.5 days of data taking, with ellipse fitting performed for each hour based on the prominence of 0.7 Hz peak after coherent subtraction of motion data from witness sensors.}
    \label{fig:ellipse_parameters_evolution}
\end{figure}

The variation of ellipse parameters between subsequent fits is shown in Fig.~\ref{fig:ellipse_parameters_evolution}. It can be seen that the variation is within at most 20\% for the radius ratio $R_1/R_2$, with the value trending downward over the measurement time, while there is a significant variation in the rotation angle $\theta$ without a clear trend. It should be noted that with typical radius ratio values of around 1.1, the rotation angle is poorly constrained, i.e. large changes in the rotation angle produce only small changes in the appearance of the ellipse. However including the rotation angle in the fit improved the overall results shown in Fig.~\ref{fig:pasd_fit}.

\subsection{Comparison of techniques}
\begin{figure}
    \centering
    \includegraphics[width=.48\textwidth]{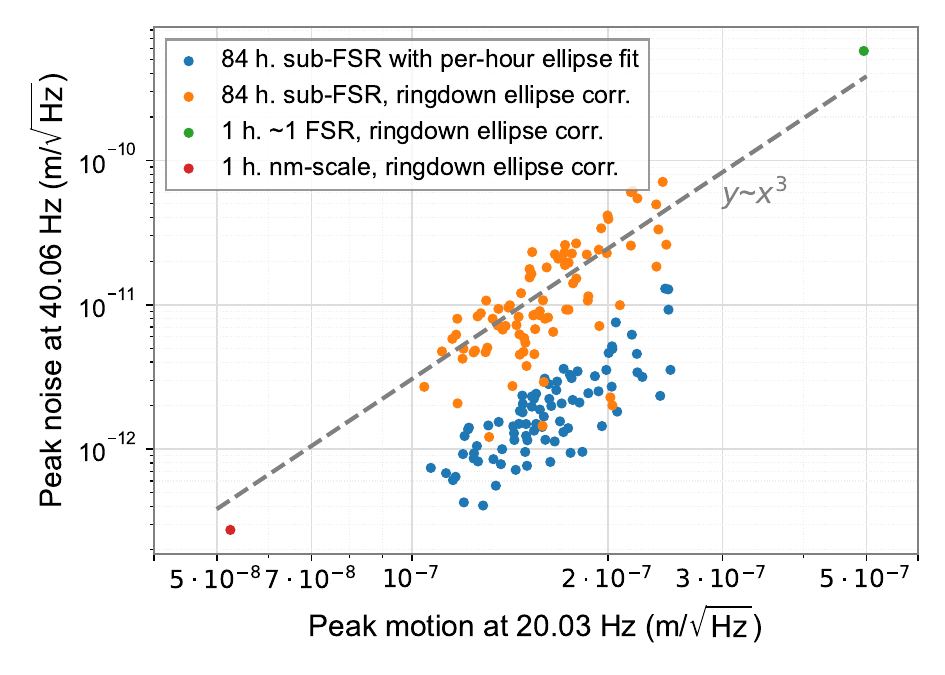}
    \caption{Noise due to residual nonlinearity for different motion regimes, calculated as the maximum value at 40.06\,Hz in one-hour ASD of HoQI distance measurement channel with subtracted witness sensor's signals, plotted against maximum motion calculated as the maximum value at 20.03\,Hz in one-hour ASD of HoQI distance measurement channel.}
    \label{fig:nice_plot}
\end{figure}

We quantified residual nonlinearity in different motion regimes by comparing ASDs of residual noise, showing the effect of ellipse correction and additional fits in post-processing. The visual prominence of the 0.7\,Hz peak quantifies the magnitude of residual nonlinearity, as it is a beat note of mechanical resonance at a frequency of around 20.7\,Hz, and oscillator's resonance frequency of 20.03\,Hz, and therefore it should vanish in an ideal readout. The same effect also applies to harmonics of the resonance frequency, one of which is around 40.06\,Hz. Another way to quantify residual nonlinearity is by plotting the noise at this harmonic versus motion at the oscillator's resonance frequency, which dominates the motion. The noise levels for different motion regimes are shown in Fig.~\ref{fig:nice_plot}. It can be observed that so measured, residual nonlinearity follows a power-law trend (close to power of 3) depending on the range of motion. The effect of per-hour ellipse fitting for 0.1-$\SI{0.3}{\micro\meter}$-range motion follows the same power-law trend but at a reduced value, consistent throughout the measurement.

\section{Summary and outlook}

In this paper, we have performed measurements of the oscillation of a 20\,Hz fused silica resonator with homodyne quadrature interferometers (HoQIs) in different regimes of motion, with amplitudes ranging from a few nanometres to a few micrometres. Nonlinearities in the readout intrinsic to HoQIs limit the sensitivity in wide frequency range in these measurements.

We have shown that the initially nonlinear signal of a HoQI can be corrected in live data by compensating mostly static ellipse errors in the quadrature signals. To measure a full ellipse and find parameters for the correction, a ringdown measurement was performed. If the motion detected by the HoQI is close to a full FSR, this method has the potential to be implemented continuously to correct for possible drifts in the ideal parameters. This would further increase the calibration robustness e.g. of HoQI-based seismometers, which are placed directly on the ground.

For motion of a lower amplitude on the scale of 0.1-$\SI{0.3}{\micro\meter}$, we successfully replaced the geometrical ellipse fitting with one that evaluates and minimises the amplitude of noise peaks arising from nonlinearities in post-processing. While possible in principle, some optimisation would be necessary to implement this approach as a continuously updating correction of live data. 

It could also be shown that whitening filters are compatible with HoQI signals and further increase their sensitivity. If the filter response is not characterised well enough, whitening filters can be an additional source of nonlinearities. Depending on the progress of live-correction methods, the required accuracy or stability of this filter characterisation could potentially be relaxed in the future. 

The signals with larger motion in this experiment on the scale of one FSR and above (around $\SI{0.5}{\micro\meter}$ range) showed a small hysteresis, which could not be corrected by the first order ellipse fitting. It could be shown that this effect is repeatable and can be suppressed in principle by separately fitting a higher order function to each direction of motion. While not compatible with the resonator used in this experiment, there is an alternative HoQI design using retroreflectors instead of mirrors which makes it less sensitive to changing alignment. If the hysteresis is really caused by small periodic misalignments of the sensor, the retroreflector HoQI should be affected much less by it. 

Overall, our methods substantially increase the applications of HoQIs to cover a much larger range of motion and still achieve the required noise floors. These methods further show that HoQIs are a good candidate for the proposed upgrades to the LIGO detectors' seismic isolation and for use in future gravitational wave detector designs. 

\section*{Acknowledgements}

Authors JL, AB, SAL, MC, GC, FK, HL, PS, ST, JvW, DSW acknowledge funding by the Deutsche Forschungsgemeinschaft (DFG, German Research Foundation) under Germany’s Excellence Strategy – EXC-2123 QuantumFrontiers – 390837967.
JJC and PB acknowledge funding in the framework of the Max-Planck-Fraunhofer cooperation project ``Glass technologies for the {Einstein} Telescope (GT4ET)''.
The authors would like to thank David Hoyland from the University of Birmingham for providing schematics for the HoQI readout electronics. 
We are also grateful for many helpful discussions and advice from Conor Mow-Lowry, Alexandra Mitchell, and Gerhard Heinzel.

\bibliography{bibliography.bib}

\end{document}